\documentclass[twocolumn,prl,amsmath]{revtex4}
\usepackage{graphicx}
\begin{document}
\title {Phase transitions in a network with range dependent 
connection probability}
\author{Parongama Sen}

\affiliation{Department of Physics, University of Calcutta,
92 A.P. C. Road, Calcutta 700009, India.}
\email {parongama@vsnl.net, paro@cubmb.ernet.in}
\author {Kinjal Banerjee and Turbasu Biswas }
\affiliation{ Department of  Physics, Presidency College, College Street, Calcutta 700073,
India.}
\begin{abstract}
We consider  a  one-dimensional network in which the 
  nodes at Euclidean distance $l$  can have long range  
connections  
with a probabilty  $P(l) \sim l^{-\delta}$ in addition to nearest neighbour
connections. This system has been shown to exhibit small world behaviour 
for $\delta < 2$ above which its behaviour is like a regular
lattice. From the study of the  clustering coefficients, we 
show that there is a transition to a random network 
at $\delta = 1$. The finite size scaling analysis
of the  clustering coefficients obtained from 
numerical simulations indicate that
a continuous phase transition occurs at this point.  
Using these results, we find  that the  two transitions occurring in this 
network can be detected in any dimension 
by the behaviour of a single
quantity, the average bond length.  
The phase transitions in all dimensions are non-trivial in nature.

\end{abstract}

\maketitle


A  network is a collection of nodes  which  are connected either 
directly or indirectly    
by links. There are  two extreme examples of networks:
the regular and the random network.
In a regular network (with infinitely many nodes), the
probability that any two arbitrary nodes are connected is 
vanishingly small
while  in random networks, this probability
remains finite. 
The two main properties which distinguish these two networks  are the chemical
distance and the clustering coefficient.  The
chemical distance is the average shortest distance between any two nodes and 
is a long distance property.
In a network with $L$ nodes, the chemical distance typically behaves as
$S_L \sim \ln L $ when it is random,  while $S_L \sim L^{1/d}$ in a regular
network  in $d$ dimensions. 
The  clustering coefficient ${\cal C}$ is the average fraction of connected triplets.  
Since the clustering coefficient  measures the local connectivity 
structure it is a short range property.
Typically ${\cal C}$  is high for the regular
network and low for the random network. 

Recently,
another kind of network, the small world network \cite{Watts1}, was proposed 
which shows
random network-like  properties at  large scales and regular network-like 
properties at small scales.
Precisely, the chemical distance $S_L$ behaves as $  \ln (L)$ while 
${\cal C}$  assumes a high value (comparable to a regular
network) in this network.
Small-world effect  can be developed  out of   a  regular lattice
having  local connections when  
long range links or connections are allowed 
to exist even with a very small  probability.

The underlying structure of a wide range of
networks including social, biological, internet and transport networks
has been argued to be small world like 
\cite{bararev}. 
Additional random long
range connections 
in model systems like Ising chains or percolation networks 
also lead to new critical behaviour \cite{Ising,newman,perco}.
Many of these networks also show scale-free behaviour, i.e., 
if $Q(k)$ is the number of
 nodes having $k$ connections, then $Q(k) \sim k^{-\gamma}$ in a scale free network.

 In the  Watts-Strogatz (WS) model \cite{Watts}, 
 the nodes are arranged in a ring.
Small world behaviour is observed when  nearest neighbour links are rewired
randomly with a probability $p$.
Later it was shown that
there is a continuous phase transition occurring at $p \rightarrow 0$ \cite{newman}
from a regular to a small world phase. 
For all values of  ~$0 < p < 1$,  the network remains small world-like with 
$S_L   \sim \ln (L)$ and a high value of ${\cal C}$.
Only at  $p = 1$, the network
behaves like a random one when 
 ${\cal C}$ vanishes.
The transitions in the WS model, which is a standard prototype model
for small world behaviour, are therefore trivial as the critical points do not 
separate phases of different critical behaviour.  
In a critical   network with  nontrivial
phase transitions   the 
critical points
  should have different phases on either side 
such that the   small world phase  emerges as a truly
intermediate phase in between the random and regular phases.
\begin{figure}
\includegraphics[clip, width=8cm]{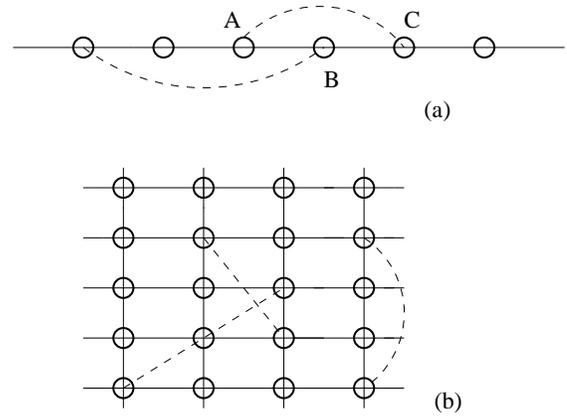}

\caption{ A section of the network under consideration  in 
(a) one and (b) two dimensions:  the solid lines denote the
 nearest neighbours links between the nodes (circles). The dashed lines 
denote the long range connections occurring with the 
probability given in (\ref{eq1}). In (a), the nodes ABC form a triplet cluster.}
\end{figure}

In this communication, we show the existence of  a  network in which such 
 non-trivial transitions can be achieved 
 by tuning an appropriate parameter. 
 In this network  \cite{blumen,klein,psbkc}, in  
addition to the nearest 
neighbour links, connection to a node at an Euclidean  distance $l$ 
is  present with a probability 
\begin{equation}
\label{eq1}
P(l) \sim l^{-\delta}.
\end{equation}
Such a network can model the behaviour of linear polymers in which
connections to further neighbours indeed occur with a power law probability.
However, in order to simulate the polymer properly,
additional constraints have to be considered \cite{psbkc}. 
Studies on the physical layout of the Internet also strongly
suggest that the connections are governed by a power law probability \cite{Internet}.

This network shows small world effect  below   $\delta = 2$ for the
one dimensional case \cite{blumen,psbkc} while 
according to \cite{klein}, there is evidence of small world behaviour at
$\delta = d$
for the $d$-dimensional network.

We have shown  that  there are two transitions 
occurring in such a network in one dimension:\\
(i) A random to small world network transition at $\delta = 1$\\
(ii) A small world to regular network transition at $\delta = 2$.\\
The second transition is already known to exist
 \cite{blumen,psbkc} although the nature of the transition
 has not been explored in detail. 
 We focus our attention on the first
 transition which has not been investigated before in either this or any other 
 network to the best of our knowledge.
 We have been able to locate the critical point  and  the 
 characteristic critical exponents
for this transition.

 The other important result of our study is the
 identification of a single quantity which can detect both the
 transitions. This analysis  can be   extended to higher dimensions
 and  the  critical points are  detected in any dimension $d$. 
 The phase diagram
 in the $\delta-d$ plane
 shows that the phase transitions are non-trivial in all dimensions.

First we  examine the properties of this network 
 in detail in one dimension.
The one-dimensional network consists of $L$ nodes occupying the sites
of a one dimensional lattice of length $L$ such that the
nearest neighbour links are of unit length
(see Fig. 1).
In addition to the nearest neighbour links there are links between nodes at distance $l$ with the 
probability given by (\ref{eq1}). Note that in one dimension, the
Euclidean distance between two nodes coincides with the number of 
nearest neighbour links 
separating  them. 
The  nearest neighbours 
links are always present in this network. This
corresponds to the situation in many realistic networks like 
linear polymers, Ising models etc.
When the probability is normalised one has
\begin{equation}
\Sigma_{ 1 < l \le L} P(l) = 1.
\end{equation}
The above normalisation condition essentially implies that the
average number of long distance connections for each node is one.
This enables a restriction on the network as  the
number of long distance bonds is  conserved. 
If the value of $\delta$
is made very high, most of the long distance connections 
will be restricted to the near neighbours and one will effectively get a 
model with short range connections only.
As $\delta$ is made smaller,
further neighbour bonds  will be
chosen.
 For very small values of  $\delta $,
connections to nodes at all possible distances are made and the network
behaves like  a random network.

From the above picture and the knowledge of the existence of a 
non-trivial phase transition 
from small world to regular network-like behaviour occurring at $\delta = 2$,  
 we expect that all three kinds of
behaviour, regular, random and small world  will be present in this model -
or in other words, there will be three regions along the $\delta$ axis:
$\delta < \delta_c^{(1)}$ where it behaves like  a random network,
$ \delta_c^{(1)}  < \delta < \delta_c^{(2)}$ where it is small world like
and $\delta > \delta_c^{(2)}$ where it is like a regular network.
A non-zero value of  $\delta_c^{(1)} $ will signify a non-trivial
transition.
In order to study the  transition from the random to 
the small world phase  it is suffcient to   
study the clustering coefficients (the chemical distance has similar
scaling behaviour in both phases).
Here we consider clusters which are triplets with three members A, B and C.  The
condition that they form a cluster is that if B is connected to A and C,
there is also a connection between A and C.
Since the nearest neighbours are always present, we 
classify the possible clusters in three classes:

1. Clusters with two nearest neighbour (nn) links each of length unity   
and one next nearest neighbour  link  of length 2.

2. Clusters with one nn link, the other two links are
of length $l_1> 1$ and $ l_1 +1 $. 

3. Clusters in which  there is no nn link;   the 
links have lengths $l_1> 1$,  $l_2> 1$, and  $l_1 + l_2 $. 

 Note that the triangular inequality of the link lengths 
 is not valid in one dimension as the distances are always measured along the
 chain.


Let ${\cal C}_i$ be the probability of the occurrence 
of a cluster belonging to the $i$th class ($i = 1,2,3$).
In the continuum limit when $l$ varies continuously, ${\cal C}_i$
take the forms:

\begin{subequations}
\begin{eqnarray}
{\cal C}_1& =& \frac {2^{-\delta}}{\int P(l) dl}\\
{\cal C}_2& = &\frac {\int_{2}^{L-2}  P(l_1)P(l_1+1) dl_1 }{({\int P(l) dl})^2}\\
{\cal C}_3& =& \frac {\int_{l_1,l_2}  P(l_1)P(l_2)P(l_1+l_2) dl_1 dl_2 }{({\int P(l) dl})^3}
\end{eqnarray}
\end{subequations}
In the last equation, the integration variables $l_1,  l_2$
satisfy 
$l_1 + l_2 < L-2$.

${\cal C}  = \Sigma ^{3}_{i=1}{\cal C}_i$ is then the clustering coefficient of the
system. 
We find that for $\delta  < 1$, ${\cal C}_i $ vanishes for
all $i$ as $L \rightarrow \infty$. In particular,  to the leading order, 
${\cal C}_1
\sim L^{\delta -1}$ while ${\cal C}_2$ and ${\cal C}_3$  
are $O(L^{-1})$. 
The vanishing of ${\cal C}$ below $\delta = 1$ indicates that
in this region the network is random. 
For $\delta  > 1$, all the three
quantities remain finite in the same limit.
Hence  ${\cal C} $ may be interpreted as  an order parameter like quantity which vanishes
at $\delta = 1$ where the transition from random to small world phase 
takes place. 

In order to find out the nature of the   transition occurring  at $\delta = 1$ 
(which we identify as 
$\delta_c^{(1)}$),
we perform numerical simulations and find out the clustering coefficient
for  chains of different lengths  with long range  connections existing with
a probability given by (\ref{eq1}). 

\begin{center}
\begin{figure}
\vskip -2cm
\includegraphics[clip, width=12cm]{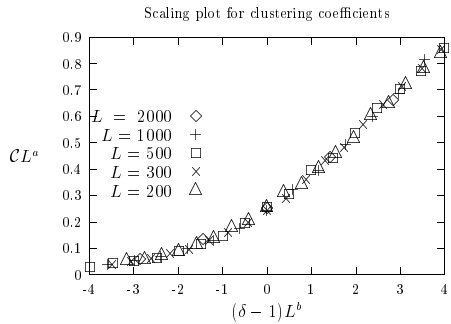}
\vskip -9cm
\caption{The data collapse of the scaled  clustering coefficients are
shown for one dimensional chains of different lengths $L$ with
periodic boundary conditions. Here $a = 0.228$ and $b= 0.258$}
\end{figure}
\end{center}

In thermodynamic system, the finite size scaling for a quantity $\Phi$ is 
given by 
\begin{equation}
\Phi(t)= L^{\phi/\nu}f(t L^{1/\nu})
\end{equation}
where $t$ is the deviation from the 
critical point, $\phi$ is the critical exponent associated with 
$\Phi $ ($\Phi \sim t^{-\phi}$)
and $\nu$ is the correlation length exponent.
We use a finite size scaling form for the 
clustering coefficients which is  analogous to that used in
thermodynamic phase transitions:
\begin{equation}
\label{scaleG}
{\cal C} = L^{-a}g((\delta_c^{(1)} -\delta)L^{b}).
\end{equation}
Using  $\delta _c^{(1)} =1$, the value obtained  in   the
continuum case,  we  
find a very good data collapse 
when ${\cal C} L^a$ is plotted
against $(\delta -1)L^b$  with the values 
$a = 0.228 \pm  0.017$ and $b = 0.258 \pm 0.030$ (see Fig. 2).
These values  are obtained using the Bhattacharjee-Seno method of 
data collapse \cite{somen}.
For large values of $x$, it is expected that $g(x) \sim x^{a/b}$.
Note that the value of $a/b$ is the estimate of the exponent $\beta$ as we have
interpreted ${\cal C} $ as the "order parameter"  
(i.e., ${\cal C}  \sim (\delta-\delta_c^{(1)})^\beta$ for $L \rightarrow \infty$), 
therefore  $\beta = 0.89 \pm 0.04$. 
Also, the 
finite size scaling form indicates that $\nu = 1/b = 3.87 \pm 0.51$.

From the above, we conclude that there is a continuous phase transition
occurring at $\delta_c^{(1)} =1$
with characteristic  exponents $\nu \sim 3.87 $ and $\beta = 0.89$.
Here $\beta$ describes how the clustering coefficient vanishes
as one approaches the random network and  the exponent $\nu$ is associated with 
a diverging length scale.
Hence in this network the small world phase appears as an 
intermediate phase between
 the random and regular phases and the characterisric behaviour of the
 network can be controlled 
 by tuning the parameter $\delta$. 
\begin{center}
\begin{figure} 
\vskip -2cm
\includegraphics[clip, width = 10cm]{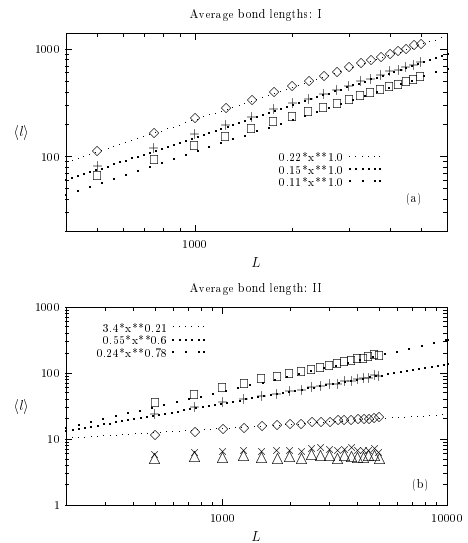}
\vskip -4.0cm

\caption{The scaling behaviour of the  average bond length
$\langle l \rangle$ for one dimensional chains of different lengths $L$
is shown. In (a), $\langle l \rangle$ varies as $L$ for $\delta < 1$. The
curves are drawn for $\delta = 0.2, 0.6$ and $0.8$ from top to bottom.
In (b) the best fit lines
for the curves for $1 < \delta < 2$ show agreement with  $ \langle l \rangle \sim 
	L^{2-\delta}$. The data are shown for $\delta = 1.2, 1.4, 1.8,  2.2 $ and $2.4$
	from top to bottom.
	For values of $\delta > 2$, $\langle l \rangle$
	does not depend on $L$.}
	\end{figure}
\end{center}

In general, the clustering coefficient and the chemical distances can detect 
either of the
two phase transitions occurring in a network.  
 This 
is because they have similar scaling behaviour in two of the 
phases and a different behaviour in the third. Interestingly,
in the present model, we find a quantity, the average 
bond length, which shows  different scaling behaviour in each of these
three phases.  In the continuum limit, 
 the average bond length $\langle l\rangle$ 
shows the following scaling behaviour:
\begin{subequations}
\begin{eqnarray}
\langle l \rangle  &\sim &L  ~~~ ({\delta < 1})\\
  &\sim &L^{2-\delta}    ~~~({ 1 < \delta < 2})\\
  &\sim & O(1)  ~~~ ({  \delta > 2}).
\end{eqnarray}
\end{subequations}
We immediately notice that the 
crossovers occur at $\delta_c^{(1)}$ and  $\delta_c^{(2)}$.
Hence we find that this is a key quantity 
since both the transitions can be detected 
from it.
This quantity is   also simple to calculate.
The numerical values of $\langle l \rangle $ for  discrete lattices agree with the above results as
shown in Fig. 3.

Although the transition points can be  located from
the behaviour of $\langle l \rangle$, estimating   the exponents is not
straightforward 
as it is difficult to cast the behaviour of $\langle l \rangle$ in 
a standard finite size scaling form as in (\ref{scaleG}).
\begin{figure}
\includegraphics[clip, width = 6cm]{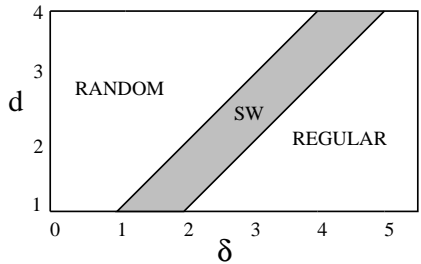}

\caption{The phase diagram of the network in the $\delta-d$ plane.
SW denotes the small world phase.}
\end{figure}

The one dimensional network which we have discussed so far
can be generalised to any dimension $d$ where the nodes occupy
the sites of a $d$-dimensional hypercube of length $L^{1/d}$ (see Fig 1b).
Each node is connected to its  $2d$ nearest neighbours and
long range links to further neighbours occur with a probability given
in (\ref{eq1}). (Note that the structure of this  network is different 
from  that of a linear polymer embedded in a $d$ dimensional
lattice although the long range bonds occur with a power law probability 
in both systems.)
For $d > 1$, neither $\delta_c^{(1)}$ nor $\delta_c^{(2)}$ is known.
In principle, these can be estimated by studying the chemical distance and the
clustering coefficients  ias in one dimension but the calculations 
become much more complicated.
The average bond length  $\langle l \rangle$, however, can be easily 
calculated
in any general dimension and we find $\langle l \rangle$
again shows  different kinds of behaviour in the three regions 
$ 0 < \delta < d ~~~({\rm{where}}~~ \langle l \rangle \sim L),  d < \delta < d+1~~ 
({\rm{where}}~~\langle l \rangle \sim L^{d-\delta + 1}), 
~~{\rm{and}} ~~\delta > d+1~~ 
({\rm{where}}~~ \langle l \rangle \sim O(1))$.
From the results of one dimension we proceed to claim 
that the transition points in $d$ dimensions are $\delta_c^{(1)} = d$ and $\delta_c^{(2)} = d+1$. 
This indicates that the small world region again exists as an
intermediate phase between a growing  random region
and a regular region. The width of the 
small world phase is independent of the dimensionality. 
The phase diagram in the $\delta-d$ plane  is shown  in Fig. 4. 

As mentioned earlier, many real networks exhibit scale-free 
behaviour which implies that the degree distribution 
is a power law. Here we checked that there is no scale free
behaviour in any regime in the one dimensional case. One can expect a scale
free behaviour  when the network shows small
world effect. The absence of scale free behaviour confirms that
a small world network is not necessarily scale-free.
On the other hand, when distance dependence in the form of 
(\ref{eq1}) is introduced in a growing network, several interesting features 
are observed \cite{mannaps}. 

In summary, our analysis of a model network where the additional long
range bonds are present with a probability dependent on the 
Euclidean distance separating the nodes shows that
there is a continuous phase transition occurring at
a finite value of the parameter $\delta$ where the
clustering coefficient behaves like an order parameter.
The transition separates  a random and a small world 
phase. With evidence of a transition from a small world to regular behaviour
already existing, we find that this network
can be tuned to show  regular, random and small world behaviour for different
values of $\delta$. 
The different behaviour occurring on the two sides of
the critical points  mark the existence of non-trivial phase
transitions. This is a feature absent in the familiar  WS model 
of small world network, 
where the transitions  are reminiscent of the zero temperature 
phase transitions occurring in the one dimensional Ising model.

Comparing the results obtained from the clustering coefficients and
the chemical distances, we find that 
both the transitions can be detected from 
the behaviour of the average bond lengths. 
This analysis can be extended to any dimension 
and the transition points located. We believe that
this idea could be useful in general 
for locating critical points in network whenever  
the bond length is a meaningful 
quantity.


Acknowledgements: 
PS acknowledges DST (India) grant no. SP/S2/M-11/99.


\end{document}